\documentclass[aps,twocolumn,prd,superscriptaddress,nofootinbib,floatfix]{revtex4-2}

\usepackage{mathtools}
\usepackage{amsfonts}
\usepackage{amssymb}
\usepackage{mathrsfs}
\usepackage{bbm}
\usepackage{slashed}
\usepackage{amsmath}
\usepackage{bm}
\usepackage{bbold}
\usepackage{physics} 

\usepackage{graphicx}
\usepackage{color}
\usepackage{colortbl}
\usepackage{array}

\usepackage{float}
\usepackage{placeins}
\usepackage{booktabs}
\usepackage[caption=false]{subfig}
\usepackage{makecell}
\usepackage{tabstackengine}

\usepackage{xspace}
\usepackage{hyperref}
\usepackage[nameinlink]{cleveref}
\usepackage{bookmark}
\usepackage{siunitx}

\usepackage{xifthen}
\usepackage{xcolor}
\hypersetup{
	colorlinks,
	linkcolor={red!75!black},
	citecolor={blue!75!black},
	urlcolor={blue!75!black}
}

\usepackage[utf8]{inputenc}
\allowdisplaybreaks[4]

\newcommand{\mub}{\mu_{{B}}}

\newcommand{\feyn}[1]{
	\setbox0=\hbox{\ensuremath{#1}}
	\hbox to\wd0{\hbox to0pt{\hbox to\wd0{\hss/\hss}\hss}\box0}}

\renewcommand{\c}{\; ,}
\newcommand{\p}{\; .}
\newcommand{\idp}{\int\!\frac{d^3p}{(2\pi)^3\!} }

\def\Eq#1{\Cref{#1}}
\def\Fig#1{\Cref{#1}} 

\setkeys{Gin}{width=0.48\textwidth}
\graphicspath{{./figures/}}

\newcommand{\gettitle}{}

\newcommand{\getGiessenAffiliation}{\affiliation{Institut f\"ur Theoretische Physik, Justus-Liebig-Universit\"at Gie\ss en, 35392 Gie\ss en, Germany}}

\newcommand{\getHFHFAffiliation}{\affiliation{Helmholtz Research Academy Hesse for FAIR (HFHF), Campus Giessen, Giessen, Germany}}

\hypersetup{
	colorlinks,
	linkcolor={red!75!black},
	citecolor={blue!75!black},
	urlcolor={blue!75!black},
	pdftitle={\gettitle},
	pdfauthor={Zelle},
	pdfkeywords={effective theory} {analytic continuation}
	{correlations functions} {hadronization}
	{functional renormalisation group}
	{real time} {bound states}
	bookmarksopen=true,
	bookmarksopenlevel=2,
	bookmarksnumbered=true
}

\begin{document}

\title{Imprints of the nuclear liquid-gas phase transition on net-baryon number fluctuations}

\author{Mattia Recchi}
\getGiessenAffiliation

\author{Shi Yin}
\email{shiyin.dalian@gmail.com}
\getGiessenAffiliation
\getHFHFAffiliation

\begin{abstract} 

We investigate net-baryon number fluctuations in the high-density, low-temperature region of the QCD phase diagram using the parity-doublet model (PDM) under the mean-field approximation. We compute the fluctuation ratios up to sixth order around the nuclear liquid-gas (LG) phase transition, where the high-order ratios are particularly sensitive. To connect the results with heavy-ion experiments, we test several different scenarios of chemical freeze-out. We find that near the LG transition the extracted fluctuations depend strongly on the choice of freeze-out curve. We self-consistently determine four freeze-out points from preliminary results of the STAR Collaboration. Comparing the experimental data with the PDM results along these four points, we find that the model describes the low energy ($\sqrt{s_{NN}}\lesssim$ 4 GeV) data well. This suggests that nucleon interactions and the LG phase transition may contribute significantly to the fluctuations in low energy heavy-ion collisions.

\end{abstract}
	
\maketitle

\section{Introduction}

The low density region of the QCD phase diagram is well described by chiral symmetry breaking and the quark confinement mechanism \cite{Aarts:2023vsf,Fischer:2026vkc,Fischer:2026uni}. As the temperature decreases, the hot strongly interacting matter created after the Big Bang undergoes the chiral and deconfinement transitions, thereby converting quark matter into hadronic matter. High energy heavy-ion collision experiments help us explore this process and, after the chemical freeze-out of particles, provide various observables from which the details of the phase transition can be inferred. As the chemical potential increases, the fluctuations induced by the density make the phase structure more complex. The chiral crossover gradually sharpens and may eventually terminate at a critical end point (CEP) \cite{Luo:2017faz,Fu:2019hdw,Pandav:2022xxx}. More exotic phases may also be excited as the baryon density increases \cite{Pawlowski:2025jpg,Pisarski:2021qof,Fu:2024rto}. In the low-temperature region of the phase diagram near the baryon saturation density, nuclear matter also exhibits the nuclear liquid-gas (LG) phase transition \cite{Sauer:1976zzf,Borderie:2019fii,Elliott:2013pna}, which is unrelated to the quark-gluon plasma (QGP) and is governed entirely by nucleon interactions.

Fluctuations of conserved charges, especially the net-baryon number fluctuations, are considered one of the most promising heavy-ion observables for detecting signals of the CEP \cite{Luo:2017faz,Stephanov:2008qz,Stephanov:2011pb}. The phase I and II Beam Energy Scan (BES) program carried out by the STAR Collaboration at the Relativistic Heavy Ion Collider (RHIC) has measured net-proton number fluctuations in heavy-ion collisions at energies between 7.7 and 200 GeV \cite{Adam:2020unf,STAR:2025zdq,STAR:2025ydk}. The fixed-target (FXT) experiments from STAR \cite{STAR:2022vlo,QM2025sweger} together with HADES Collaboration \cite{HADES:2020wpc} have also produced results of the fluctuations at lower collision energies. Combining the two sets of experimental data, one may infer that a possible signal of the CEP could lie between 4 and 7.7 GeV, a range consistent with recent theoretical predictions.

The first-principles QCD predictions for the location of the CEP are mainly given by functional methods. Current calculations using the functional renormalization group (fRG) \cite{Fu:2019hdw,Fu:2026qnl,Wang:2026xwa,Pawlowski:2025jpg} and the Dyson-Schwinger equations (DSE) \cite{Gao:2020qsj,Gao:2020fbl,Gunkel:2021oya} consistently constrain the possible location of the CEP to the region of baryon chemical potential between 600 and 700 MeV. These two methods have also been used to compute net-baryon number fluctuations and to interpret and predict the observables measured in heavy-ion experiments, see, e.g., \cite{Fu:2021oaw,Fu:2023lcm,Lu:2025cls,Lu:2026ezr,Isserstedt:2019pgx}. The predicted collision energy range in which the CEP signal may appear in the baryon number fluctuations is consistent with the estimate from the experiments discussed above.

When entering the high-density region generated by the FXT experiments, the situation becomes considerably more complex. Many non-critical fluctuations begin to have a non-negligible impact on the heavy-ion observables, such as finite acceptance effect \cite{Zhang:2019lqz}, global baryon number conservation \cite{Braun-Munzinger:2020jbk,Fu:2023lcm,Vovchenko:2020tsr,Vovchenko:2021yen,Poberezhniuk:2026bfv,Li:2023kja,Li:2026tcq}, baryon scattering and transport \cite{Chen:2024zry,Lin:2026bso}, and non-equilibrium evolution effects \cite{Berdnikov:1999ph,Mukherjee:2016kyu,Tan:2025bsv}. As the collision energy decreases, the production of QGP gradually diminishes, and the effects of nucleon-nucleon interactions in the system gradually increases. As the density increases, the chemical freeze-out curve of heavy-ion collisions gradually bends toward the low-temperature region and moves away from the chiral phase transition boundary \cite{Fu:2021oaw,Andronic:2017pug,Lu:2026ezr}. When the freeze-out curve dives into the low-temperature region, the nucleon interaction and even the critical fluctuations of the nuclear LG phase transition can influence the observables measured along it.

The hadron resonance gas (HRG) model is an important tool for studying the thermodynamic properties of the hadron gas based on hadronic degrees of freedom \cite{Karsch:2003zq,Huovinen:2009yb,Vovchenko:2019pjl}. By including interactions, e.g., van der Waals interactions and excluded volume effects, the HRG model can produce the nuclear LG phase transition. Within this framework, baryon number fluctuations have been studied, see, e.g., \cite{Vovchenko:2016rkn,Poberezhnyuk:2019pxs,Vovchenko:2017cbu,Samanta:2017yhh}. There are also studies of baryon number fluctuations based on models in which the nucleon interactions are mediated by mesons, such as the parity doublet model (PDM) \cite{Koch:2023oez,Mukherjee:2016nhb,Marczenko:2025kpv}, the Walecka model \cite{Yang:2023gfy,Shao:2020lsv} and also other similar types of model \cite{Marczenko:2017huu,Fukushima:2014lfa}. In these models, the baryon number fluctuations near the LG phase transition exhibit similar behavior. 

However, for a meaningful comparison with heavy-ion data, the determination of the chemical freeze-out curve is equally important, because near the LG phase transition even a slight change in the shape of the freeze-out curve can lead to substantially different behavior of the baryon number fluctuations. The commonly used chemical freeze-out curve is obtained by fitting the experimental data compiled in \cite{Andronic:2017pug}. However, the existing experimental data are mainly in the low density region, with very few data points at high density, in particular near the LG phase transition. Moreover, the functional form that the freeze-out curve should take is unknown, so the fluctuation calculations in this region carry relatively large uncertainties. In studies of baryon number fluctuations using functional methods, the chemical freeze-out temperature and chemical potential are determined by using the lower order fluctuations measured in heavy-ion experiments, see, e.g., \cite{Fu:2015amv,Lu:2026ezr}. Although this method may not yield a universal freeze-out curve, it can, to some extent, eliminate the model dependence within calculations based on the same model. In this work, we employ the PDM in the mean-field approximation to compute baryon number fluctuations of various orders, and, combining it with the freeze-out extraction method mentioned above, we discuss the influence of different curves on the fluctuations. 

This paper is organized as follows. In \Cref{sec:fluc}, we briefly introduce the definition of the baryon number fluctuations and the method of calculation. In \Cref{sec:PDM}, we review the PDM employed in this work. In \Cref{sec:result}, we present our results: \Cref{subsec:fluc_finitemub} introduces the results at zero and finite chemical potential, \Cref{subsec:fluc_phase} presents the results over the entire phase diagram and discusses the choice of different chemical freeze-out curves, and \Cref{subsec:sqrt_fluc} presents the fluctuations evaluated along the freeze-out curves and compares them with the fRG calculations and heavy-ion experimental data. Finally, in \Cref{sec:summary}, we summarize and discuss our results.

\section{Baryon number fluctuations}
\label{sec:fluc} 

As observables for detecting the signal of CEP, baryon number fluctuations are widely computed and investigated. From the perspective of statistical physics, the high-order fluctuations of particle number distributions can be obtained through derivatives of the potential with respect to the corresponding chemical potentials under the grand canonical ensemble. So the particle number fluctuations can be calculated through the pressure of the system
%
\begin{align}\label{eq:chi}
\chi^B_n=\frac{\partial^n}{\partial(\mu_B/T)^n}\frac{p}{T^4}\,.
\end{align}
%
Although the effect of global charge conservation cannot be neglected in heavy-ion experiments at high density, in this work we focus only on the effect of the LG phase transition on the observables. Therefore we still adopt the grand canonical ensemble (GCE) in our calculations. The pressure can be obtained by the grand thermodynamic potential with 
%
\begin{align}\label{eq:p_omega}
p=-\Omega[T,\mu_B]\,.
\end{align}
%
The potential is a function of temperature $T$ and baryon chemical potential $\mu_B$. The relations between the 1st to 6th order cumulants and the particle number fluctuations can be given by
%
\begin{align}
    \chi^B_1&=\frac{1}{VT^3}\langle N_B\rangle\,,\\[2ex]
    \chi^B_2&=\frac{1}{VT^3}\langle (\delta N_B)^2\rangle\,,\\[2ex]
    \chi^B_3&=\frac{1}{VT^3}\langle (\delta N_B)^3\rangle\,,\\[2ex]
    \chi^B_4&=\frac{1}{VT^3}\bigg(\langle (\delta N_B)^4\rangle-3\langle (\delta N_B)^2\rangle^2\bigg)\,,\\[2ex]
    \chi^B_5&=\frac{1}{VT^3}\bigg(\langle(\delta N_B)^5\rangle-10\langle(\delta N_B)^2\rangle\langle(\delta N_B)^3\rangle\bigg)\,,\\[2ex]
    \chi^B_6&=\frac{1}{VT^3}\bigg(\langle(\delta N_B)^6\rangle-15\langle(\delta N_B)^4\rangle\langle(\delta N_B)^2\rangle\\[2ex]
    &-10\langle(\delta N_B)^3\rangle^2+30\langle(\delta N_B)^2\rangle^3\bigg)\,.
\end{align}
%
The angle brackets denote the ensemble average. The fluctuation of the baryon number around its average is given by $\delta N_B=N_B-\langle N_B\rangle$. The higher order relations are also introduced in, e.g., \cite{Fu:2021oaw,Luo:2017faz}.
Note that there are volume factors for every order of the fluctuation. When we want to compare our results with the experimental data, we can instead use the ratios of the cumulants to eliminate the volume effect, e.g.,
%
\begin{align}
R^B_{n,m}=\frac{\chi^B_n}{\chi^B_m}\,.
\end{align}
%
Current heavy-ion collision experiments focus on probing signatures of the CEP through the non-monotonic behavior of the kurtosis $R^B_{42}$ of the net-baryon number distributions as a function of collision energy, while higher-order fluctuations have also been measured. Although such measurements of higher-order fluctuations require very high precision, the existing experimental data already provide some qualitative understanding. In this work, we focus on the region of chemical potential higher than that of the CEP, and employ the PDM to study the possible effects of the nuclear LG phase transition on high-order baryon number fluctuations at high density. In the next section, we briefly introduce the PDM and how to use it to compute the fluctuations.
\section{Parity-Doublet Model}
\label{sec:PDM}

The PDM is a generalization of the linear sigma model \cite{Gell-Mann:1960mvl} in which the sigma meson and the pion fields are coupled to the nucleon $N(939)$ and its negative-parity partner, here assumed to be the $N^*(1535)$ resonance. In the PDM, baryons interact through meson exchange, which allows for a good description of the thermodynamics in the high-density, low-temperature region of the phase diagram where the LG phase transition occurs. The computation in this work is based on the setup from \cite{Recchi:2025pyy}.

In this work, we consider two sets of spin-$1/2$ baryons, $N_1$ and $N_2$, transforming in the two-flavor chiral representations $(\tfrac{1}{2},0)\oplus(0,\tfrac{1}{2})$. These fields are coupled in a chirally invariant way to the $O(4)$ meson multiplet $(\sigma,\vec\pi)$. Imposing the “mirror assignment” \cite{Jido:1998av}, in which the left-handed component of $N_2$ transforms as the right-handed component of $N_1$ and vice versa under $SU(2)_L\times SU(2)_R$, allows the introduction of a chirally invariant mass term $m_0$. The chiral invariant mass term can be conceptually interpreted as the gluonic field contribution to the mass of the baryons. This mass term is shared by the two parity partners and it has no reason to be affected by chiral symmetry and its spontaneous breaking \cite{Yang:2018nqn}.

The baryonic and baryon-meson interaction part of the Euclidean Lagrangian $\mathcal{L}_F$ can then be written as
\begin{equation}\label{eq:L_f}
\begin{split}
\mathcal{L}_F &=\bar{N}_1\Big(\slashed\partial + g_1(\sigma + i\gamma_5 \vec\tau \cdot \vec\pi)\Big)N_1 \\[2ex]
&+ \bar{N}_2\Big(\slashed\partial + g_2(\sigma - i\gamma_5 \vec\tau \cdot \vec\pi)\Big)N_2 \\[2ex]
&+ m_0 \Big(\bar{N}_1 \gamma_5 N_2 - \bar{N}_2 \gamma_5 N_1\Big)\,.
\end{split}
\end{equation}
Here $g_1$ and $g_2$ denote the Yukawa couplings, while the opposite signs in the pion–baryon interactions reflect the mirror structure of the chiral transformations of $N_1$ and $N_2$. When chiral symmetry is spontaneously broken, the $\sigma$ field acquires a vacuum expectation value $\langle \sigma \rangle = f_\pi \simeq 93 \ \mathrm{MeV}$, generating Dirac mass contributions. Throughout this work we assume homogeneous condensates and, without loss of generality, choose the chiral condensate to lie in the $\sigma$ direction, so we have $\langle\pi\rangle=0$ by parity and isospin symmetry.

It is possible to diagonalize the Lagrangian in the physical basis of $N_+=N(939)$ and $N_-=N^*(1535)$ obtaining:
\begin{equation}
\label{eq:free_fermions}
\mathcal{L}_F =\bar{N}_+(\slashed\partial + m_+)N_+
+ \bar{N}_-(\slashed\partial + m_-)N_-  \,,
\end{equation}
with the physical masses $m_\pm$ given by 
\begin{equation}
\label{eq:masses}
m_\pm =\frac{1}{2} \sqrt{(g_1+g_2)^2\sigma^2 + 4m_0^2} \pm \frac{g_1-g_2}{2}\sigma \, .
\end{equation}
To fix the parameters $g_1$ and $g_2$ in the vacuum, we can use the masses of the baryons $m_N\equiv m_+=939$~MeV for the nucleon and $m_{N^*}\equiv m_-=1510$~MeV for the $N^*(1535)$ as its parity partner \cite{ParticleDataGroup:2024cfk}, together with $\expval{\sigma} = f_\pi$. As chiral symmetry gets restored in the medium, the sigma-field diminishes, when chiral symmetry is fully restored, i.e., $\langle\sigma\rangle=0$, according to \Cref{eq:masses}, the parity-partner baryons become degenerate in mass with finite $m_0$. 

As shown in \cite{Recchi:2025pyy}, it is possible to write a Renormalization Group (RG) invariant parametrization that allows for a convenient inclusion of renormalized vacuum contributions. We follow the same derivation and use the same parameters in the previous work.
The mesonic part of the Lagrangian reads
\begin{equation}\label{eq:U}
    \begin{split}
        \mathcal{L}_{\mathrm{mes}} &= \sum_{n=1}^4 \frac{1}{2n} \frac{a_{2n}}{m_0^{2n-4}} \Big(m_+^{2n} + m_-^{2n}\Big) - c\sigma \p
    \end{split}
\end{equation}
No derivative of the mesonic fields is included due to the isotropy and homogeneity hypothesis in the MF approximation. The term $-c\sigma$, with $c=f_\pi m_\pi^2$ provides the explicit breaking of the chiral symmetry, with $m_\pi=138$ MeV at vacuum.

To provide the hard-core repulsion necessary to properly describe the nuclear interaction we include the usual vector field $\omega$. This is implemented via the following replacements: 
\begin{align}
	\slashed{\partial} &\to \slashed{\partial} - ig_\omega \gamma^\mu \omega_\mu \c\\[2ex]
    \mub &\to \tilde{\mu}_B =  \mub - g_\omega \omega_0  \c \\[2ex]
    \mathcal{L}_\mathrm{mes}  &\to  \mathcal{L}_\mathrm{mes}   - \frac{1}{2}m_\omega \omega_0^2 \p   
\end{align}
Using standard thermal field theory techniques and including the renormalization of the sea contribution we obtain the grand potential $\Omega = \Omega_\mathrm{fer} + \Omega_\mathrm{mes} + \Omega_\mathrm{vac}$ with

\begin{align}
\Omega_\mathrm{ferm} &= 2T \sum_\pm \idp \Bigg\{ \ln \bigg[1-f\qty(\frac{E_\pm -\tilde\mu}{T})\bigg] \nonumber\\[2ex]
&\qquad\qquad\qquad \quad + \ln \bigg[1-f\qty(\frac{E_\pm +\tilde\mu}{T})\bigg] \Bigg\} \c \\[2ex]
\Omega_\mathrm{mes} &= \mathcal{L}_\mathrm{mes} \c\\[2ex]
\Omega_\mathrm{vac} &= \frac{1}{\pi^2} \sum_\pm \frac{m_\pm^4}{4}\ln \big(m_\pm/{m_0}\big) \p
\end{align}
$f(x) = {1}/{(e^x+1)}$ is the Fermi-Dirac distribution.
The renormalization scale in the vacuum contribution $\Omega_\mathrm{vac}$ is arbitrary fixed to $m_0$, considered the natural scale of this model. The expectation values of the fields are obtained by extremizing $\Omega$ with respect to the mesonic field, i.e., solving the equation of motion (EoM)
\begin{equation}
	\pdv{\Omega}{\sigma} = 0 \; , \quad
	\pdv{\Omega}{\omega} = 0 \; , \quad
\end{equation}
at fixed $T$ and $\mu_B$. The corresponding solutions of the EoM are $\sigma_0$ and $\omega_0$. This allows for the computation of the pressure \Cref{eq:p_omega} and its derivatives used in the calculation of baryon number fluctuations $\chi^B_n$ in this work.
%
\begin{figure}[t]
\includegraphics[width=0.48\textwidth]{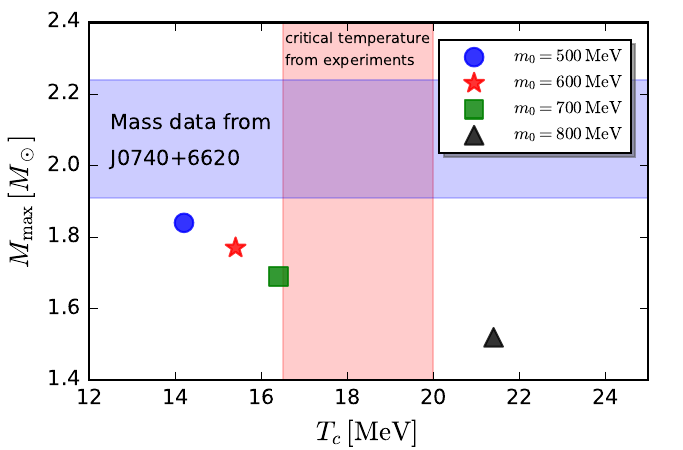}
\caption{The predictions of the PDM on maximum mass and critical temperature of the LG phase transition for different values of $m_0$. The red-shaded band is obtained from the central values of the empirical measurements in \cite{Karnaukhov:2008be} while the blue-shaded band is from the 95\% CL value of the mass of J0740+6620 \cite{Dittmann:2024mbo}. Our parameter choice $m_0=600$ MeV is the closest to the intersection of the shaded regions}
\label{fig:benchmark}
\end{figure}
%
The coefficient of \Cref{eq:U} as well as the vector coupling $g_\omega$ are obtained by fitting the properties of nuclear matter at saturation density. The Yukawa couplings and the explicit chiral symmetry breaking strength can all be determined by the nucleon masses and pion mass. Then the chiral invariant mass $m_0$ is the only parameter left. The value of $m_0$ can be determined by comparing the calculated results with observables, i.e., Mass-Radius (M-R)-relations of neutron stars and the location of the LG critical point \cite{Karnaukhov:2008be}. In \Cref{fig:benchmark} we give the maximum mass supported by a neutron star ($M$) and the critical temperature of the LG transition ($T_c$) predicted by the model consistently with each choice of $m_0$, see \cite{Recchi:2025pyy}. In the figure, an empirical estimation for $T_c$ ($M$) is also highlighted with a red-shaded (blue-shaded) background \cite{Karnaukhov:2008be}. Since it is clear that $m_0=600$ MeV is the closest point to the intersection of the shadowed regions, we choose this value to obtain results that are in reasonable agreement with both observations. In the present work, in order to keep the overall model simple, the constraint we impose on the value of $m_0$ is relatively coarse. For more studies on the value of $m_0$ and its constraint from the neutron star physics, see Refs. \cite{Minamikawa:2023eky,Gao:2024chh,Gao:2025vdc,Gao:2025nkg}.

Having fixed the parameters of the PDM, we now discuss in the next section the numerical results of baryon number fluctuations at large baryon chemical potential within this framework.
\section{Numerical Results}
\label{sec:result}
In this section, we discuss the results from the PDM. In \Cref{subsec:fluc_finitemub}, we present the baryon number fluctuations calculated within the PDM at large baryon chemical potential. In \Cref{subsec:fluc_phase}, we discuss the fluctuations on the phase diagram and several different methods to determine the chemical freeze-out parameters at high chemical potential. Finally in \Cref{subsec:sqrt_fluc}, we present the PDM calculations across the entire phase diagram and extract the fluctuations along the chemical freeze-out curve. Unlike previous studies that focused on the qualitative structure of fluctuations near the LG transition, we systematically investigate how the choice of freeze-out curve affects the extracted fluctuations, and directly confront the results with the latest STAR FXT and HADES data as a function of collision energy.
\subsection{Baryon number fluctuations at finite density}
\label{subsec:fluc_finitemub}
With the grand canonical potential given by the PDM at finite temperature and density, we can directly use the \Eq{eq:chi} to compute the baryon number fluctuations of different orders. Following a procedure similar to that of \cite{Fu:2021oaw,Fu:2023lcm}, we compute the pressure at several values of the chemical potential and then evaluate its high-order derivatives by numerical differentiation.

\begin{figure}[t]
\includegraphics[width=0.48\textwidth]{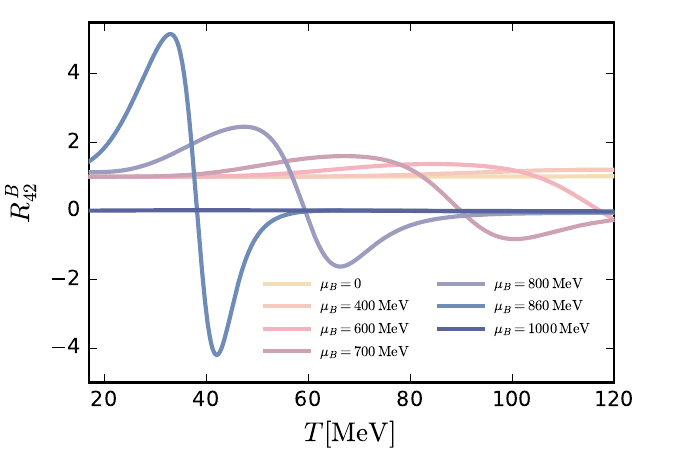}
\caption{Baryon number fluctuation $R^B_{42}$ computed within the PDM as functions of temperature from 0 to 120 MeV under several values of chemical potential between 0 and 1 GeV.}
\label{fig:fluc_PDM}
\end{figure}

In the \Fig{fig:fluc_PDM}, we choose several different values of the baryon chemical potential to present the temperature dependence of the fluctuations. The following analysis investigates the effect of the LG phase transition then the chemical potential range is chosen from zero to 1 GeV, covering different regions of the phase diagram.
%
%

We find that, for low density data, the $R^B_{42}$ remains equal to one. This result is consistent with the hadron resonance gas model \cite{Vovchenko:2016rkn,Poberezhnyuk:2019pxs,Braun-Munzinger:2003pwq} and lattice QCD \cite{Bazavov:2020bjn,Bazavov:2017dus,Bazavov:2017tot,Borsanyi:2018grb} results at low temperature. This suggests that the net-baryon number follows a Skellam distribution corresponding to an independent Poisson distribution of baryons and antibaryons. This indicates that the model is in the dilute gas limit at low temperature and small chemical potential. With the increase of the chemical potential, $R^B_{42}$ gradually begins to show fluctuations, which become increasingly pronounced. Even if the real LG transition only extends in a relatively small region of the phase diagram for $T>T_c$, we can still identify a residual crossover transition, for instance in the chiral condensate magnitude. The position of the most rapid variation area between the peak and valley of the fluctuations corresponds to the pseudo-critical temperature ($T_{\mathrm{pc}}$) of the LG phase transition at the given chemical potential. It can be seen that the $T_{\mathrm{pc}}$ decreases as the chemical potential increases. Finally, when the chemical potential or temperature exceeds the LG phase transition region, $R^B_{42}$ tends toward zero. High order fluctuations are suppressed by the nucleon-meson interactions, leading to the vanishing of the kurtosis. These behaviors are similar to the results from the Polyakov-loop extended quark-meson model, e.g., \cite{Fu:2015naa,Fu:2021oaw}.

\subsection{Baryon number fluctuations on the phase diagram and chemical freeze-out}
\label{subsec:fluc_phase}
%
\begin{figure*}[t]
\includegraphics[width=0.9\textwidth]{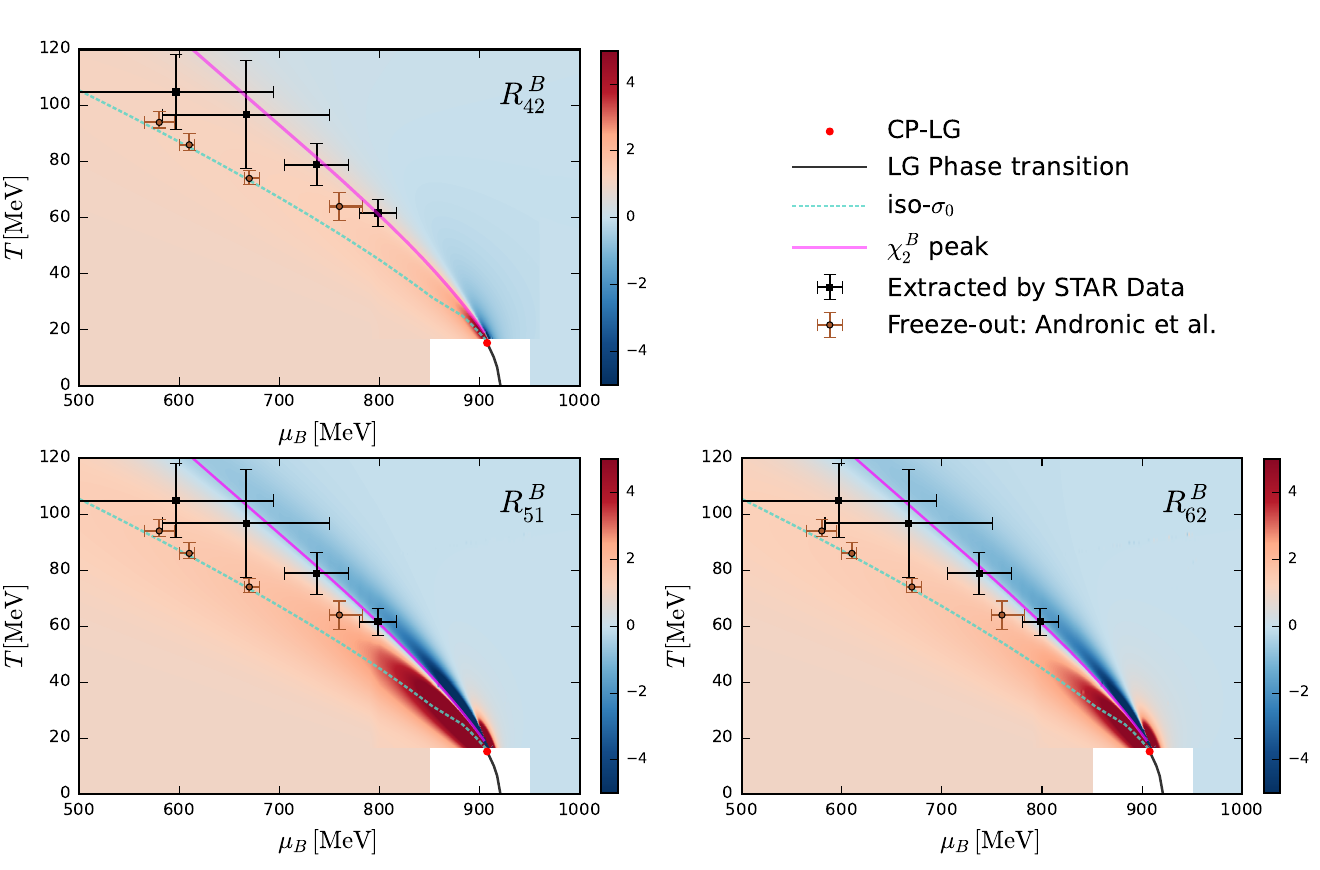}
\caption{The upper-left, lower-left, and lower-right panels show heatmaps of $R^B_{42}$, $R^B_{51}$ and $R^B_{62}$, respectively, displaying the results for the fluctuations on the phase diagram in the vicinity of the LG phase transition. The red dot marks the LG critical point, and the black solid line the first-order LG phase transition line. The teal dashed line is a contour of the chiral condensate, along which $\sigma_0=88$ MeV. The magenta solid line indicates the peak position of the second-order fluctuation $\chi^B_2$. The black squares with error bars denote the freeze-out parameters extracted by the preliminary $R^p_{21}$ and $R^p_{31}$ data from STAR collaboration \cite{Sweger:2025}. The brown circles with error bars denote the freeze-out parameters from \cite{Andronic:2017pug}.}
\label{fig:phasePDM_all}
\end{figure*}
%
We perform the computation of the PDM on the whole phase diagram and study the behavior of the $R^B_{42}$, $R^B_{51}$ and $R^B_{62}$ around the LG critical point. Since we ultimately aim to express our results as a function of the collision energy in order to compare them with experimental observables, in this section we also discuss the determination of the freeze-out parameters in the vicinity of the LG phase transition.

In \Fig{fig:phasePDM_all}, we show the results of the baryon number fluctuation $R^B_{42}$, $R^B_{51}$ and $R^B_{62}$ on the phase diagram. The fluctuation data in this figure cover a chemical potential range from 500 MeV to 1 GeV, and a temperature range from 1 MeV to 120 MeV. Note that, we do not show the data below a temperature of 17 MeV around the LG phase transition line. Because in the calculation of the fluctuations, we employ a numerical differentiation method to take the derivative of the grand thermodynamic potential with respect to the chemical potential. Therefore, in the regions close to the critical point (CP) and the first-order phase transition, the chemical potential interval used for differentiation must be extremely small, which leads to significantly large numerical errors in these regions. Meanwhile, the region of our interest also lies at temperatures higher than that of the CP, so we do not need the data around the phase transition line in this work. In the figure, warm colors represent positive values, while cool colors represent negative values. 

As can be seen from the three panels, the lower-order and higher-order fluctuations exhibit similar behavior, they vary rather smoothly at low chemical potential and become sharper as the CP is approached. However, near the CP the higher-order fluctuations develop more complex structures than the lower-order ones, in agreement with the calculation in \cite{Fu:2023lcm}. For the kurtosis $R^B_{42}$, it clearly separates into positive and negative regions near the CP, with the boundary between the two regions given by the crossover of the LG phase transition. The structures of $R^B_{51}$ and $R^B_{62}$ are rather similar to each other, both exhibit a positive–negative–positive "sandwich" structure near the CP. The fluctuation ratios of all orders approach 1 in the low-temperature, low-chemical potential region, and tend to 0 in the high temperature, high-density region.

This complex structure gives rise to a problem. In order to compare with experimental data, we extract the values of fluctuations along the chemical freeze-out curve. Due to the novel structure of the fluctuations, the exact position of the freeze-out curve on the phase diagram can strongly affect the extracted fluctuation results. Because of the limited experimental freeze-out data points available at high chemical potential, especially near the LG critical point, as well as the complexity of multiple effects, e.g., non-equilibrium effects, baryon number conservation effects, and so on, in heavy-ion experiments in the high-density region, the freeze-out parameters ($T$ and $\mu_B$) in this region have not yet reached a reliable conclusion. Therefore, we should first determine the possible chemical freeze-out curve parameters in the high-density region. 

In this work, we adopt several different scenarios for the freeze-out curve, which we introduce separately below. First, in \Fig{fig:phasePDM_all} we show the contour of the chiral condensate in teal dashed line. On this line, the value of the chiral condensate is $\sigma_0=88$ MeV. As discussed in \cite{Recchi:2025pyy}, we choose this value because this contour agrees very well with the four freeze-out data points from \cite{Andronic:2017pug}, i.e., the brown error-bars on the phase diagram. This is also found in \cite{Floerchinger:2012xd}. Therefore, here we list the chiral condensate contour as one of the candidates for the chemical freeze-out curve. In addition, we also calculated the peak positions of the second order fluctuation $\chi^B_2$, which corresponds to the magenta curve in \Fig{fig:phasePDM_all}. We can see that the magenta line lies right between the red area and blue area of the heat map. This indicates the crossover line of the LG phase transition and finally ends at the critical point.

Finally, we also use the preliminary STAR FXT data point to extract the freeze-out points. Note that the HADES Collaboration has also reported a data point at low energy \cite{HADES:2020wpc}. since their measurement uses a different rapidity range from that of STAR, we use only the STAR data points to extract the freeze-out curve, in order to maintain consistency. The extraction method was proposed in \cite{Lu:2026ezr}. This method utilizes low-order baryon number fluctuations to constrain the chemical freeze-out parameters, which are then used to predict the behavior of higher order fluctuations. The freeze-out parameters determined by this method can partially eliminate the model dependence of the calculations, thereby providing relatively reliable predictions of higher order fluctuations. Similar to \cite{Lu:2026ezr}, we use $R^B_{21}$ and $R^B_{31}$ to jointly constrain the freeze-out parameters. Since the experimental data of low-order fluctuations have smaller uncertainties, the parameters can be determined relatively accurately. In \Fig{fig:phasePDM_all}, we use black squares with error-bars to show the points extracted by the low-order data. The uncertainties in the chemical potential and temperature directions are introduced from the error-bars of the experimental data during the extraction of the freeze-out parameters. The extraction procedure can be found in \Cref{app:extract}. It can be seen that in the high-density area, e.g., the 3 GeV data point, the uncertainties are the smallest, while at lower chemical potentials the uncertainties are larger. This happens because the lower order fluctuations vary relatively smoothly at low chemical potential so the range of values within the error bars covers a comparatively large region of the phase diagram. At high chemical potential the fluctuations change rapidly, so that the region of the phase diagram covered by the error bars becomes much smaller.

We now summarize the three freeze-out extraction methods introduced above:
%
\begin{itemize}
    \item Constant chiral condensate\,,
    \item Peak position of $\chi^B_2$\,,
    \item Extracted by low-order cumulants\,.
\end{itemize}
%

Now that we have established the conditions to compare the calculated results with the experimental data, in the next subsection, we can extract the fluctuations along the different scenarios of chemical freeze-out curves and analyze the validity of each freeze-out curve.

\subsection{Fluctuations on the freeze-out curves}
\label{subsec:sqrt_fluc}
%
\begin{figure*}[t]
\includegraphics[width=0.9\textwidth]{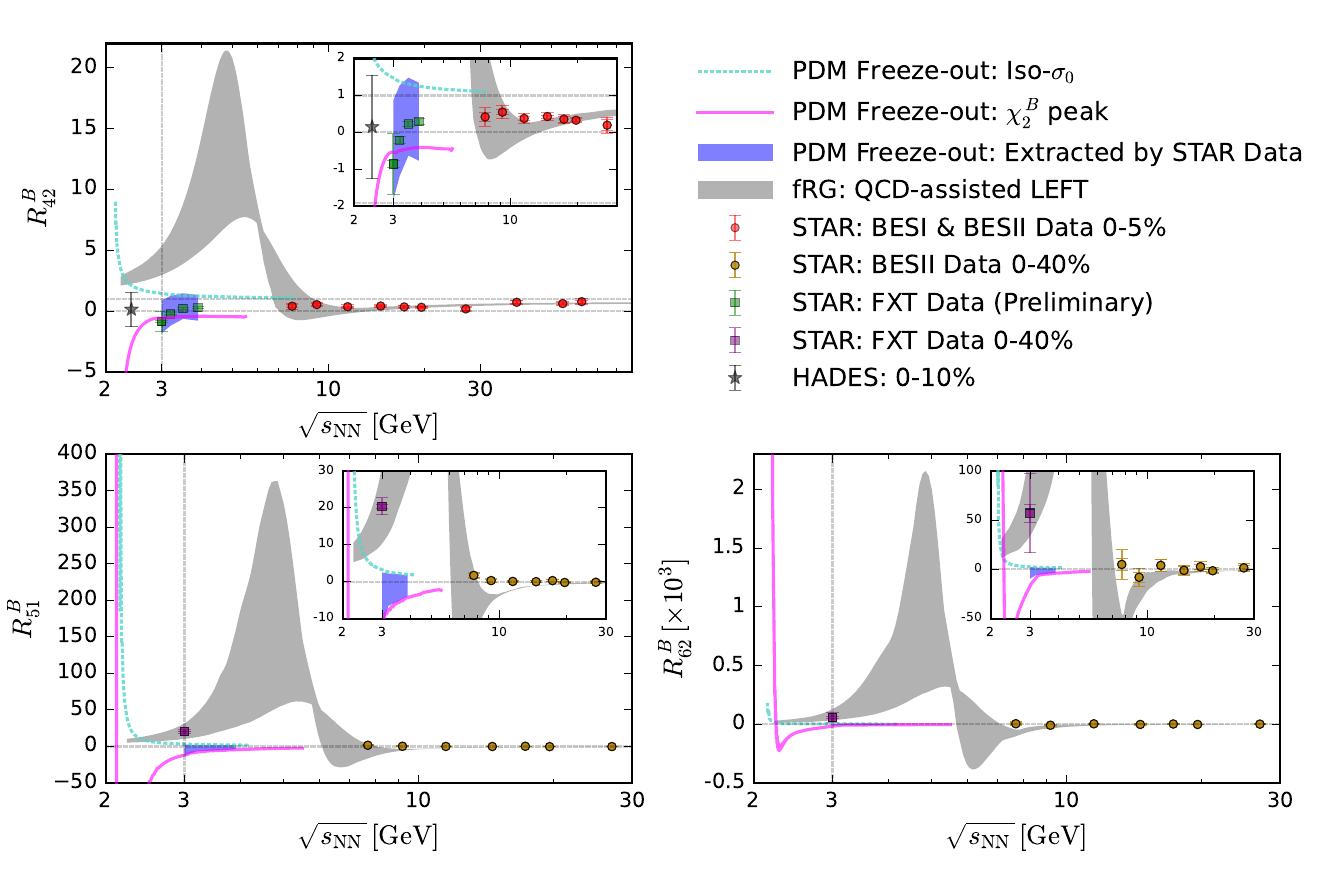}
\caption{$R^B_{42}$, $R^B_{51}$ and $R^B_{62}$ of the baryon number fluctuations evaluated along the chemical freeze-out curves, together with the net-proton fluctuation data from BESI, BESII and FXT experiments. The teal dashed line shows the fluctuations extracted along the chiral condensate contour $\sigma_0=88$ MeV in the PDM, and the magenta lines show the fluctuations extracted along the curve of the $\chi^B_2$ peak in the PDM. The blue bands represent the PDM results evaluated along the freeze-out curve extracted from the preliminary FXT data. The gray bands show the fRG QCD-assisted LEFT results \cite{Fu:2023lcm}. The red error bars denote the experimental data from BESI \cite{Adam:2020unf} and BESII \cite{STAR:2025zdq} at 0–5$\%$ centrality, the yellow error bars the higher-order fluctuation data from BESII at 0–40$\%$ centrality \cite{STAR:2025ydk}, and the green and purple error bars the preliminary FXT data \cite{STAR:2022vlo,QM2025sweger}. The gray error bar is given by the HADES Collaboration at 2.4 GeV \cite{HADES:2020wpc}.}
\label{fig:sqrts_all}
\end{figure*}
%
We directly read the data of $R^B_{42}$, $R^B_{51}$ and $R^B_{62}$ on the freeze-out curves, and we use the error bars on the STAR data to associate the uncertainties with the ratios. See \Cref{app:extract} for details. In \Fig{fig:sqrts_all}, we show the fluctuation results on the freeze-out curves that are discussed in the last subsection. At the same time, we also put the fRG QCD-assisted LEFT results, the BES data and the FXT data points on the plot for comparison.

We first focus on the results of $R^B_{42}$ (upper left panel). The data points from the BES program phase I and II cover collision energies from 7.7 GeV to 200 GeV. In this collision energy range, the theoretically predicted non-monotonic behavior, i.e., the possible observational signal of the CEP \cite{Stephanov:1999zu,Stephanov:2008qz}, has not yet been observed. This collision energy range roughly corresponds to the region of baryon chemical potential below 400 MeV on the QCD phase diagram. This means that the signal of the CEP potentially exists at some point below the collision energy 7.7 GeV. 

The gray band in the plot is given by the QCD-assisted LEFT computed by fRG \cite{Fu:2023lcm}. This result is based on the Polyakov-Quark-Meson (PQM) model with the fRG-QCD data input. This model describes the chiral phase transition very well, and investigates the behavior of the high order fluctuations around the CEP of the chiral phase transition. However, even though the Polyakov loop potential captures aspects of confinement at low temperature, the absence of explicit nucleon degrees of freedom and their interactions still prevents the model from describing the high-density, low-temperature behavior of QCD compellingly. Even if the QCD-assisted LEFT can provide us with the non-monotonic behavior of $R^B_{42}$ on the freeze-out curves which would be a possible signal of the CEP, at low collision energy it becomes increasingly unreliable. For the grand canonical ensemble assumption of the model, the $R^B_{42}$ will always go to one when the freeze-out curve dives into the low-temperature region at high density. This, however, does not appear to hold for the experimental data. Due to the decrease in collision energy, the production of QGP after the collision gradually diminishes, and nucleon scattering gradually becomes more relevant. At low-temperature, the LG phase transition may become increasingly relevant and can provide an important contribution to the fluctuations in this region

The cumulant ratios $R^B_{42}$ observed by the FXT experiment below 7.7 GeV \cite{Sweger:2025,STAR:2021fge,HADES:2020wpc} are all smaller than 1, corresponding to the data points marked by the green error-bars in \Fig{fig:sqrts_all}. As is mentioned in the introduction, it is well known that under the conditions of low energy heavy-ion collisions, many non-critical effects, non-equilibrium effect, global charge conservation effect, etc., can influence the final observed results. One point this work aims to highlight is that the fluctuations arising from the nuclear LG phase transition at high density may also be one of the contributing factors responsible for $R^B_{42}$ being below unity. For $R^B_{51}$ and $R^B_{62}$ at 3 GeV, the data still lie within the gray band, but they correspond to 0–40$\%$ centrality. If we consider a narrower centrality of 0–5$\%$, they would likely deviate from the fRG result as well \cite{Xu:2026qvi}.

We introduce the results given by the PDM. As explained in last subsection, the freeze-out curve in the high-density region lacks experimental constraints, so the two curves in \Fig{fig:phasePDM_all} lie on opposite sides of the LG phase boundary, thus yielding completely different behaviors of $R^B_{42}$. The teal dashed line is given by the contour of the chiral condensate $\sigma_0=88$ MeV. Since the value is still close to the vacuum value $f_\pi$, the model is in the chiral symmetry broken phase. The $R^B_{42}$ on the condensate contour is always a positive value and increases monotonically from unity as the collision energy decreases. $R^B_{51}$ and $R^B_{62}$ exhibit the same behavior. The other freeze-out curve hypothesis is determined by the peak positions of $\chi^B_2$, which lies essentially on the phase boundary of the LG crossover and slightly shifted toward the chiral restoration region. The results on this line are given by the magenta lines. We see that $R^B_{42}$ along this curve remains negative throughout and decreases monotonically in the low collision energy region. These two curves represent two extreme possibilities: if the freeze-out curve lies in the lower temperature region, $R^B_{42}$ remains positive throughout, whereas if the freeze-out curve lies in the higher temperature region, $R^B_{42}$ is negative. Whereas $R^B_{51}$ and $R^B_{62}$ develop non-monotonic behavior in the region of smaller collision energies. This is because $R^B_{51}$ and $R^B_{62}$ have more complex structures, and $\chi^B_2$ crosses from the blue negative region to the red positive region near the transition point.

In order to reduce the model dependence to some extent and incorporate experimental data, we adopt the method from \cite{Fu:2015amv,Lu:2026ezr} to extract the freeze-out temperature and chemical potential from experimental data. The freeze-out parameters are extracted using lower order fluctuations, e.g., $R^B_{21}$ and $R^B_{31}$. These lower-order fluctuations are relatively straightforward to obtain in experimental data analysis, and the corresponding error-bars are also smaller. Through the extraction from experimental data, we obtain the black squares with error-bars in \Fig{fig:phasePDM_all} and extract the corresponding $R^B_{42}$, $R^B_{51}$ and $R^B_{62}$ values within the uncertainty boxes. Since the freeze-out relation between the temperature and chemical potential in the high-density region is unknown, fitting the extracted parameters with an arbitrary functional form can easily lead to erroneous conclusions. Therefore, we only present the results at the few energy points provided by FXT experiment. The blue error-band in \Fig{fig:sqrts_all} represents the $R^B_{42}$ extracted from the PDM results. Although the value $R^B_{42}=1$ also falls within the error of the calculation results, the PDM results decrease gradually with decreasing energy, a behavior consistent with that extracted at the $\chi^B_2$ peak. Remarkably, the $R^B_{42}$ extracted through the freeze-out parameters agrees very well with the preliminary experimental results. The HADES data point in the figure is at 2.4 GeV. This point lies right between the two freeze-out curves we adopt, which suggests that data points at even lower energies may fall near the LG phase transition and depend very sensitively on the position of the freeze-out curve in the vicinity of the transition. 

It should be noted that the model assumptions of PDM are also based on an equilibrated nucleons system in the grand canonical ensemble, and effects such as non-equilibrium and charge conservation have not been taken into account in the calculations. The reason for the good agreement with the experiment may suggest that the fluctuations from the LG phase transition at these collision energies indeed have a non-negligible effect on the whole system. We look forward to the final data release from the STAR FXT experiment, which will allow for a more rigorous analysis of the calculation results. For the higher-order fluctuations $R^B_{51}$ and $R^B_{62}$, the 3 GeV data points still lie above the PDM blue band. As they correspond only to 0–40$\%$ centrality, no firm conclusion can be drawn at this stage. However, the insets show that the blue bands follow the same decreasing trend as the fluctuations extracted along the $\chi^B_2$ peak. For all three orders, the results based on the experimentally extracted freeze-out points behave similarly to those along the $\chi^B_2$ peak. This suggests that the actual freeze-out curve may converge toward the LG critical point along the phase boundary.

\section{Summary and Conclusion}
\label{sec:summary}
In this work, we employ a PDM which is based on nucleon interactions within the mean-field approximation and compute the net-baryon number fluctuations $R^B_{21}$, $R^B_{31}$, $R^B_{42}$, $R^B_{51}$ and $R^B_{62}$ on the $T-\mu_B$ phase diagram. In the PDM, nucleons interact through meson exchange, which provides a good description of the phase structure in the high-density, low-temperature region, in particular the nuclear LG phase transition (see the discussion in \Cref{sec:PDM}). A comparison of the $R^B_{42}$ results with those of the fRG, HRG and lattice QCD shows that the reliability of the PDM gradually deteriorates at high temperature. We therefore restrict our calculations to the region below 120 MeV.

The main goal of this work is to compute the net-baryon number fluctuations in the high-density region, particularly at collision energies $\sqrt{s_{NN}}\lesssim 4$ GeV, and to study the possible effects of the nuclear LG phase transition on baryon number fluctuations in the high-density low-temperature region. To achieve this goal, we first compute the fluctuations of various orders on the phase diagram of the model; the results are discussed in \Cref{subsec:fluc_finitemub,subsec:fluc_phase}. We propose three different methods to extract the chemical freeze-out curve in the high-density, low-temperature region, and evaluate the baryon number fluctuations along these curves for comparison with experimental data. The methods for extracting the chemical freeze-out curve are discussed in \Cref{subsec:fluc_phase}.

In \Cref{subsec:sqrt_fluc}, we compare the baryon number fluctuations from the PDM along the chemical freeze-out curves, with the experimental data of the STAR Collaboration and HADES Collaboration as well as with the fRG results. We find that the PDM results agree better with the experimental data in the low collision energy region, e.g., $\sqrt{s_{NN}}\lesssim 4$ GeV. In particular, the freeze-out curve extracted from the lower-order fluctuations is more consistent with the preliminary experimental $R^P_{42}$ data, while the higher-order fluctuations also yield results significantly below the fRG calculation. The good agreement between the PDM results and the current experimental data suggests that nucleon interactions and the LG phase transition may provide an important contribution to the observed fluctuations in the low-energy region. In this region, calculations based solely on quark and meson degrees of freedom may become insufficient, and nucleon interactions become a non-negligible effect.

In summary, this work employs the PDM, based on equilibrium thermodynamics and the grand-canonical ensemble to study baryon number fluctuations, aiming to point out that the effects of nucleon interactions and of the LG phase transition may play an important role in low-energy heavy-ion collision experiments. At the present stage, our calculations and analysis do not yet include non-equilibrium effects, baryon number conservation or the distinction between protons and neutrons. Therefore, our results cannot rule out corrections from these non-critical effects to the baryon number fluctuations in the low-energy region. Future analyses of heavy-ion signals based on this model will be progressively refined and will incorporate a more complete treatment of non-critical fluctuations.

\section{Data availability}
The data used to produce the figures in this paper are available from the authors upon reasonable request.
\section*{Acknowledgements}

The authors are grateful to Lorenz von Smekal, Jianing Li, Jan M. Pawlowski and Fabian Rennecke for valuable discussions. 
This work is supported by the Deutsche Forschungsgemeinschaft, project no.~315477589, the Collaborative Research Center CRC-TR 211, ``Strong-interaction matter under
extreme conditions.'' 

\appendix 
\crefalias{section}{appendix}   
\crefname{appendix}{Appendix}{Appendices}
\section{Baryon number fluctuation at vanishing density}
\label{app:mub0data}
%
\begin{figure}[!ht]
\includegraphics[width=0.48\textwidth]{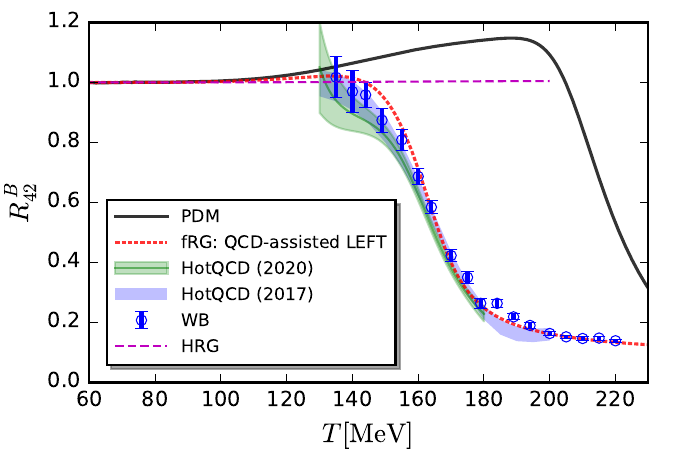}
\caption{Baryon number fluctuation $R^B_{42}$ as functions of temperature at vanishing chemical potential. The black solid line denotes the PDM result, the red dashed line gives the result of the fRG QCD-assisted LEFT \cite{Fu:2023lcm}, the purple dashed line is HRG result \cite{Braun-Munzinger:2003pwq}. The green and blue bands represent the lattice results of the HotQCD Collaboration \cite{Bazavov:2017dus,Bazavov:2017tot,Bazavov:2020bjn}, the blue error bars show the lattice results of the WB Collaboration \cite{Borsanyi:2018grb}. It is visible that PDM results begin to deviate from other results in the region beyond 120 MeV.}
\label{fig:0mub}
\end{figure}
%
In the \Fig{fig:0mub}, we show the results of $R^B_{42}$ computed by the PDM together with the HRG, fRG and lattice results. As can be seen from the figure, the $R^B_{42}$ obtained from the fRG QCD-assisted LEFT calculation agrees better with the lattice results. The PDM result, on the other hand, starts from 1 at low temperature, rises slightly with increasing temperature, and then begins to decrease at around 200 MeV. The origin of this difference lies in the fact that the fundamental degrees of freedom in the PDM are always a restricted set of baryons. Although the model can capture the chiral phase transition to some extent at high temperature, it lacks the deconfinement transition to quark degrees of freedom, and therefore cannot correctly describe the phase transition behavior at high temperature. In addition also the contribution of the majority of resonances is missing. Since the PDM gradually loses its reliability in the high temperature region, in the following calculations we focus on the low-temperature region and discard the unreliable high temperature part. We therefore restrict our calculations to temperatures below 120 MeV.
\section{Freeze-out parameters extraction}
\label{app:extract}
%
\begin{figure*}[!ht]
\includegraphics[width=0.48\textwidth]{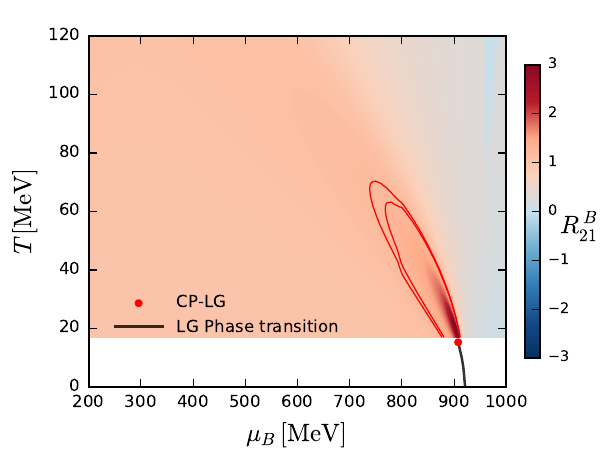}
\includegraphics[width=0.48\textwidth]{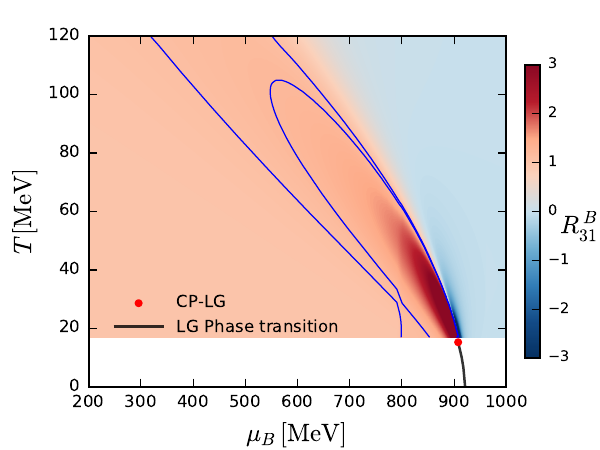}
\caption{The contours of $R^B_{21}$ (left red) and $R^B_{31}$ (right blue) correspond to the upper and lower bounds of the error-bars of the STAR FXT data at the given collision energy 3 GeV.}
\label{fig:contour1}
\end{figure*}
%
%
\begin{figure}[t]
\includegraphics[width=0.48\textwidth]{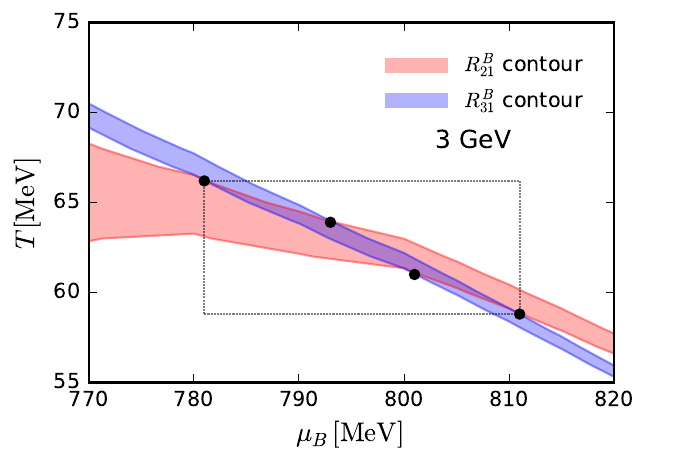}
\caption{The contours of $R^B_{21}$ (red) and $R^B_{31}$ (blue) correspond to the upper and lower bounds of the error-bars of the STAR FXT data at the given collision energy (3 GeV in this plot). The black dots correspond to the intersections of the contours, which also give the uncertainties of the extracted chemical freeze-out parameters.}
\label{fig:contour}
\end{figure}
%
In this appendix, we explain the method we use to extract the freeze-out temperature and chemical potential for the PDM. The method of using lower order fluctuations to extract freeze-out parameters was first applied to the quark-meson model in \cite{Fu:2015amv} where they used $R^B_{21}$ and $R^B_{32}$ to determine the freeze-out temperatures at different chemical potentials. Then in \cite{Lu:2026ezr}, this method was further extended to simultaneously extract both the chemical potential and the temperature from $R^B_{21}$ and $R^B_{31}$, and it was applied to DSE calculations.

In this work, we also use the contours of $R^B_{21}$ and $R^B_{31}$ to perform the extraction. The preliminary data points are given by STAR collaboration \cite{Sweger:2025,STAR:2021fge}. Here we use the data point at 3 GeV as an example. We first compute the $R^B_{21}$ and $R^B_{31}$ on the entire phase diagram, then search for the contours corresponding to the uncertainty range of the experimental data points, the intersections of these contours define the uncertainty range of the freeze-out parameters at the corresponding collision energy. 

We give the results of $R^B_{21}$ and $R^B_{31}$ on the phase diagram in \Fig{fig:contour1}. Both fluctuations vary more strongly with increasing chemical potential, becoming particularly sharp near the LG critical point. In both heatmaps, we also show contour lines of the fluctuations. The values of the two contours are extracted from the upper and lower bounds of the error-bars of the preliminary STAR results at 3 GeV. As shown in \Fig{fig:contour}, the red band is the contour of $R^B_{21}$ and the blue one is given by $R^B_{31}$. The four black dots mark the four intersection points of the contours, and the two outermost points are taken to define the uncertainty range of the freeze-out chemical potential and temperature. The same method can be applied to the other three collision energy points, thereby getting the four black squares with error-bars in \Fig{fig:phasePDM_all}.

\vfill 
\bibliography{ref-lib}%

\end{document}